\documentclass[doublecol]{epl2} 
\usepackage{amsmath}

\usepackage{graphicx}
\usepackage{dcolumn}
\usepackage{bm}

\title{Description and connection between the Oxygen order evolution and
the superconducting transition in $La_2CuO_{4+y}$}
\shorttitle{A Theory of oxygen order evolution and $T_c$ in $La_2CuO_{4+y}$ }

\author{E. V. L. de Mello }

\institute{
  \inst{1} Instituto de F\'isica, Universidade Federal Fluminense, Niter\'oi, RJ 24210-340, Brazil
}

\abstract{
A recent hallmark set of experiments by Poccia et al in cuprate superconductors 
related in a direct way, for the first time, time ordering ($t$) 
of oxygen interstitials in initially disordered  $La_2CuO_{4+y}$  
with the superconducting transition temperature $T_c(t)$. 
We provide here a description of the time ordering forming pattern domains 
and show, through the local free energy, how it affects the superconducting 
interaction. Self-consistent calculations in this granular structure
with Josephson coupling among the domains reveal that the superconducting
interaction is scaled by the local free energy and capture 
the details of $T_c(t)$. 
The accurate reproduction of these apparently disconnected phenomena establishes
routes to the important physical mechanisms involved in sample production and 
the origin of the superconductivity of cuprates.}

\begin{document}

\maketitle

\section{Introduction}

There are considerable evidences that the tendency toward phase separation
is  an universal feature of cuprate superconductors and 
others electronic oxides like manganites\cite{Muller,Dagotto}. 
The presence of hole-rich and hole poor phases were detected
in oxygenated $La_2CuO_{4+y}$ almost immediately after the discovery of the 
superconductivity in these compounds\cite{Jorgensen} and in subsequent
works\cite{Grenier,Radaelli,Wells,Campi,Lee}. Subsequent experiments have
observed evidences of complexity and electronic disorder in many other 
materials\cite{Muller,Dagotto,Bianconi}.

In order to describe this phenomenon, some theories producing phase separation
have been suggested, mainly based on doped Mott-Hubbard
insulators. Some of these theories rely on Fermi-surface nesting, which leads to a
reduced density of states or a gap at the Fermi energy\cite{Poilblanc,Machida,Zaanen,Schulz}.
Others use a competition between the tendency of
an antiferromagnetic insulator to expel doped holes and the 
long range Coulomb interaction to explain the formation of charge
ordered phases\cite{Emery,Kivelson}. Another approach suggested that
the pseudogap line $T^*$ is the onset of a first order transition, with the
development of carriers of two types, frustrated by the 
electro-neutrality condition in the presence of rigidly embedded dopant
ions\cite{Gorkov1,Gorkov2}. 

These theories describe some of the observed features, 
however they fail to predict some other basic experimental results, 
specially those related with real space inhomogeneities. Specifically, the
recent combined experiment relating the time evolution ($t$) of the domain 
growth of oxygen interstitials (i-O) in $La_2CuO_{4+y}$ 
with the subsequent measurement of the superconducting transition 
temperature $T_c(t)$\cite{Poccia} brings new possibilities that
require new approaches.

 In this letter we propose a theory that provides an
interpretation to the
two parts of the Poccia et al experiment\cite{Poccia} and gives a clear explanation
why the $T_c$ increases with the oxygen ordering.
We rely on a description of the phase separation in cuprates based 
on the time-dependent Ginzburg-Landau or Cahn-Hilliard (CH) equation 
introduced earlier\cite{Otton,Mello04}. We showed
that the CH solutions yield granular
regions where the free energy has two types of minima, for high
and low doping, and that these valleys are surrounded by steep boundaries where the 
charges can get trapped, loosing part of their kinetic energy
which enhances the mechanism of pair formation\cite{Mello09,Mello11}.
During the phase separation process the free energy barrier 
between the two (high and low density) phases varies with the time
which connects the time of oxygen ordering at high temperatures 
with the variations of the superconducting critical temperature $T_c(t)$
at low temperatures.
Scaling the superconducting interaction with the local
changes of the free energy is one of our most interesting finding.
Another new point is that in this granular-like system the resistivity 
transition temperature $T_c$ occurs when the Josephson energy 
$E_J$ among the grains is equal to $K_BT_c$\cite{Mello11}. These
new ideas are endorsed by the close agreement with the data as
described below.

\begin{figure}[ht]
    \centerline{\includegraphics[width=7.0cm]{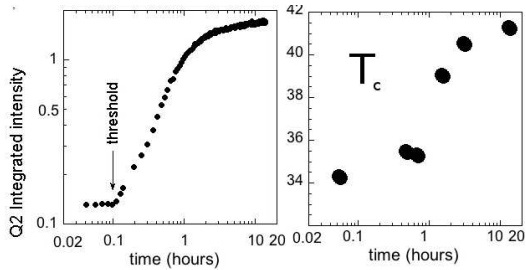}}
\caption{ Graphs taken from Poccia et al\cite{Poccia}.
On the left panel the presence of the reflected X-ray diffraction
peaks associated with the oxygen order: we see that after a threshold, 
it increases rapidly  and enter in a slow variation regime after a 
few hours. On the right panel, the corresponding measured
values of $T_c(t)$ after the system being cooled down.}
\label{Xray} 
\end{figure}

The connection between dopant atoms like the 
out of plane i-O\cite{Poccia,Fratini,Ricci} 
and the charges in the $CuO_2$ planes is well verified for 
$La_2CuO_{4+y}$, but it has also been seen in other cuprates\cite{Keren}.
It is an open question whether the electronic disorder 
is driven by the lower free energy of undoped
antiferromagnetic (AF) regions\cite{Mello09} 
(intrinsic) or by the out of plane dopant's (extrinsic origin)\cite{Keren}, 
like the i-O ordering\cite{Fratini,Poccia}. In either case, being
intrinsic or extrinsic, it is very likely 
that there is an one to one correspondence 
between the two phenomena. This one to one correspondence is a crucial
ingredient of our work and will be explored in detail here. Consequently 
the i-O ordering time evolution is assumed  to occur together
with the planar electronic phase separation (EPS) and they are
described by the same 
CH equation\cite{Otton,Mello04,Mello09,Mello11}.

\section{Phase Separation}

Below the phase separation
transition temperature $T_{ps}$, taken to be the order-disorder
transition temperature for the i-O  mobility 
at $T_m\approx 330$K\cite{Poccia}, the appropriate order parameter is the 
normalized difference 
between the local $p(i,t)$ and the average charge density $p$
$u(p,i,t)\equiv (p(i,t)-p)/p$. Here, in order to compare with 
the actual system, we perform calculations with the optimal doping $p=0.16$,
but we can perform calculations with any doping level.
Clearly $u(i,t)=0$ corresponds to the homogeneous system,
above $T_{PS}$, and  $u(i,t)=\pm 1$ corresponds to the extreme case of full
phase separation which is hardly achieved because the mobility
vanishes as the temperature goes down. In this way one expects the system
to reach an intermediated structure between homogeneous and
complete phase separated. The Ginzburg-Landau (GL) free energy functional  
is the usual $u$ power expansion,

\begin{eqnarray}
f(u)= {{\frac{1}{2}\varepsilon^2 |\nabla u|^2 +V_{GL}(u,t)}}.
\label{FE}
\end{eqnarray}
where the potential ${\it V_{GL}}(u,T)= -A^2(T)u^2/2+B^2u^4/4+...$,
$A^2(T)=\alpha(T_{PS}(p)-T)$, $\alpha$ and $B$ are constants. 
$\varepsilon$ gives the
size of the boundaries between the low and high density phases
\cite{Otton,Mello04}. The CH equation can be written\cite{Bray} 
in the form of a continuity equation of the local density of free energy $f$,
$\partial_tu=-{\bf \nabla.J}$, with the current ${\bf J}=M{\bf
\nabla}(\delta f/ \delta u)$, where $M$ is the mobility or the
charge transport coefficient that sets the phase 
separation time scale. Therefore,
\begin{eqnarray}
\frac{\partial u}{\partial t} = -M\nabla^2(\varepsilon^2\nabla^2u
- A^2(T)u+B^2u^3).
\label{CH}
\end{eqnarray}

\begin{figure}[ht]
    \begin{center}
     \centerline{\includegraphics[width=7.0cm]{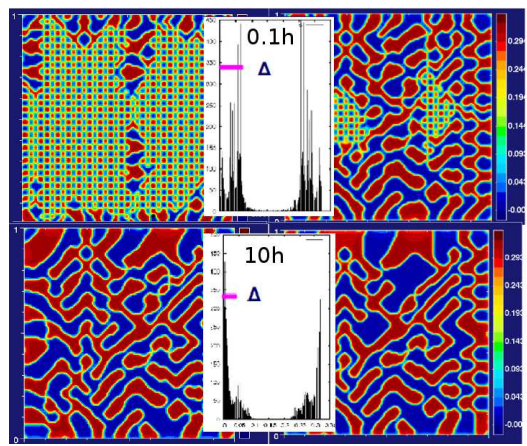}}
\caption{ (color online) Time evolution  simulation 
of the  inhomogeneous charge density with average $p=0.16$ on an array of $105 \times 105$
sites. The vertical scale gives the color code for the local densities
(red is high and blue is low density). 
Top panels are for $t=2000ts$ end $t=4000ts$. Lower panels are for the $t=20.000$
and $t=200.000ts$. In top middle the local density $p(i)$ histogram 
for the $t=2000ts\equiv 0.1h$. In the lower panel the histogram for
$t=200.000ts\equiv 10h$. The  estimated FWHM of the charge dispersion 
$\Delta$ marked in the plots  are to be compared with the experimental values.
}
\label{MapVu} 
\end{center}
\end{figure}

The phase separation process described by this CH differential 
equation occurs due to the minimization of the free energy given
by Eq.(\ref{FE}).  
The parameter $\varepsilon$ determines
the size of the grain boundaries and $A(T)/B$ the values of the order
parameter at the minima near the transition temperature.
If $A(T)$ is zero (above $T_{ps}$) there
is only one solution for the free energy and for the non-vanishing case
there are two solution corresponding to the two phases (high and
low densities)\cite{Bray}. In Fig.(\ref{MapVu}), 
we show some typical simulations of the density map  
with the two (hole-rich and hole-poor) phases given by
different colors.  The simulations were done 
with $\varepsilon=0.08$ and $A(T)/B=1$. We use a constant
value for the $A(T)/B$ ratio because usually the phase separation
occurs in a small temperature interval and the mobility ceases 
at low temperatures.

In the case of the i-O experiments\cite{Poccia} the system is heated to $T=370$K
above the mobility temperature $T_m=330$K and quenched to low temperature.
Such procedure does not yield ordered i-O peaks and has poor superconducting
order. However, by irradiating a given compound  with X-rays near T=300K, 
Poccia et al\cite{Poccia} observed, after a time threshold $t_0$,
the evolution of nucleation and growth of ordered domains accompanied
by the recovery of a robust high $T_c$ state. 

Here, we simulate the nucleation of these ordered domains 
following the EPS time evolution.
In the simulations on a  $105 \times 105$ lattice, we note that
below t=250ts (ts$\equiv$time
steps) the system has an uniform density but at t=265ts a non uniformity
with a regular checkerboard pattern sets in. The high and
low densities increase up to 2000ts when another domain pattern, more
irregular, develops as it is shown in Fig.2a. At 4000ts, the
checkerboard order rests only on less than 1/3 of the system and the
more stable irregular granular order dominates, as shown in 
Fig.2b. We make a correspondence between these ordered and disordered
structure and  that described by Fratini et al\cite{Fratini}, i.e.,
a system with two phases and two different $T_c$s. 
Above 6000ts the granular pattern dominates
and from 20000ts (Fig.2c) to 200000ts (Fig.2d) the grains grow
very slowly. To make a direct comparison with the experiments, 
we take the onset or threshold time
of the stable phase, i.e., {\it $t_0=2000$ts or 0.1h} as the beginning of
the i-O process. This connects in a simple way the oxygen ordering 
at larger  irradiated times with the raise and grow of the stable
domains shown in Fig.(\ref{MapVu}).

\begin{figure}[ht]
    \begin{center}
         \centerline{\includegraphics[width=6.0cm]{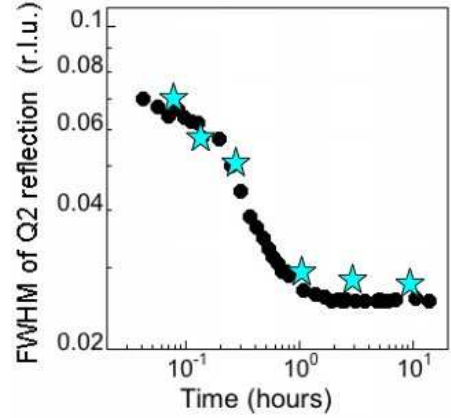}}
\caption{ (color online) 
The FWHM of the i-O X-rays diffracted dispersion peaks
\cite{Poccia} and the values of the local densities histograms
 $\Delta(t)$ calculated from 6 different times density maps
dispersion (blue stars) as shown in Fig.(\ref{MapVu} .
}
\label{FWHM} 
\end{center}
\end{figure}

In the CH approach, another way to follow a phase separation process is 
through the local densities histograms. In Fig.(\ref{MapVu}) we show this 
possibility for two selected
times and as expected, the histogram dispersion 
$\Delta(t)$ decreases as time increases. We compare directly the two dispersions,
the calculated from histograms, like those displayed in Fig.(\ref{MapVu}), 
and the experimental data in Fig.(\ref{FWHM}), {\it without any adjustable 
parameter}.

\section{The local free energy and the pair formation}
\begin{figure}[ht]
\begin{center}
    \centerline{\includegraphics[width=9.0cm]{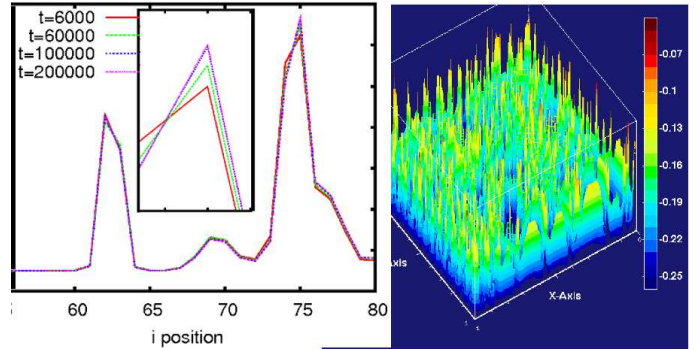}}
\caption{ (color online) A 3D view map of the potential ${\it V_{GL}}(i,t=10h)$ 
on the right panel. On the left,
the  values on 25 sites in row 
to visualize the structure of the free energy potential wells with walls 
that leads to the definition of
$V_{gb}(t)$ in the text. The values of $V_{gb}(t)$ are shown here
at $t=0.3h, 3h, 5h$ and $10h$ and are very similar. In the inset
we blow up the peaks and we see clearly a steady increase between 
$t=0.3h=6000ts$ and $t=5h=100000ts$ and the saturation above $t=5h$. 
The relative variations on $V_{gb}(t)$
are very  small but have an important role in the BdG calculations
of $\Delta_d(t)$ and on the $T_c(t)$.
}
\label{EV6200b} 
\end{center}
\end{figure}

The CH equation yields the phase separation order 
parameter $u(i,T)$, that is used to calculate the potential 
$V_{GL}(i,t)$ from free energy  (Eq.(\ref{FE})). 
This is shown by the 3D view map in Fig.(\ref{EV6200b}) for the case of
t=10h. It is also shown in the left panel
the values of  $V_{GL}(i,t)$ along 25 sites in a straight 
line in the middle of the view map of Fig.(\ref{EV6200b}) for four different
times of phase separation or X-ray irradiation. In this way, we can visualize the 
regions where the charges get trapped. The inset shows the time
variations of the barrier walls that, by assumption, scale the 
superconducting interaction\cite{Mello12}.

The calculations shown in the inset of Fig.(\ref{EV6200b}) demonstrate
how it is possible to connect the phase separation time with
the height of the $V_{GL}(i,t)$ grain boundary walls, which we define 
as $V_{gb}(t)$. 
The effect of $V_{gb}$ on the charges is 
to trap them inside the grains, and consequently they lose part of their 
kinetic energies. This loss of kinetic energy in the presence of
a two-body attractive potential favor the possibility
of Cooper pair formation and the domain walls act as a catalyst
to the superconducting state\cite{Mello12}. The  small changes on the 
size of $V_{gb}(t)$  with the time of phase separation $t$, as shown
if Fig.(\ref{EV6200b}), affects strongly the superconducting
properties. The figure shows the larger increase between $t=0.3h=6000ts$ 
and $5h=100000ts$, right before it saturates.

\section{The Superconducting Calculations}

\begin{figure}[ht]
     \centerline{\includegraphics[width=9.0cm]{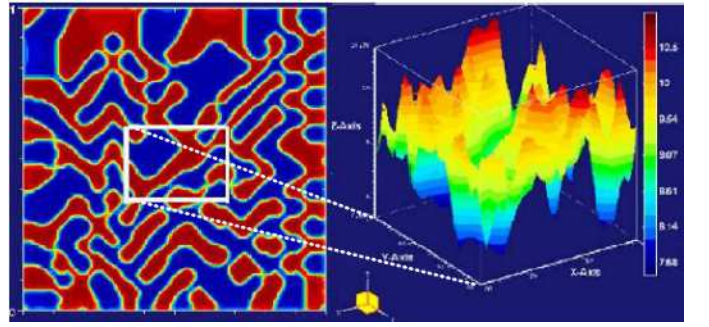}}
\caption{(color online) Left panel shows a typical density view map and the 28x28
square where the local superconducting  amplitudes $\Delta_d(i,t)$ are calculated 
by the BdG approach at $t=6000ts\equiv 0.3$h, as an example. 
}
\label{pMapDdw} 
\end{figure}

The calculated  density maps $p(i,t)$ 
on a square lattice like that of Fig.(\ref{MapVu}) are used as
the {\it initial input}  and {\it it is alway maintained fixed} through the
self-consistent Bogoliubov-deGennes (BdG) calculations. This is because
it represents the situation in which the system is quenched to low 
temperatures to measured $T_c$. We use 
nearest neighbor hopping $t_{ij}=0.15$eV and next nearest
neighbor hopping $t_1/t_{ij}=-0.27$ taken from hole doped
experimental dispersion relations\cite{Schabel}.
 For completeness, the BdG equations 
are\cite{Mello09,Mello04,DDias08}.
\begin{equation}
\begin{pmatrix} K         &      \Delta  \cr\cr
           \Delta^*    &       -K^*
\end{pmatrix}
\begin{pmatrix} u_n({\bf x_i})      \cr\cr
                v_n({\bf x_i})
\end{pmatrix}=E_n
\begin{pmatrix} u_n({\bf x_i})       \cr\cr 
                 v_n({\bf x_i}).
\end{pmatrix}
\label{matrix}
\end{equation}
These equations, defined in detail in Refs.\cite{Mello04,DDias08},
are solved self-consistently. $u_n$, $v_n$ and $E_n \ge 0$ are 
respectively the eigenvectors and eigenvalues. The 
d-wave pairing amplitudes are given by
\begin{eqnarray}
\Delta_{d}({\bf x}_i)&=&-{V_{gb}\over 2}\sum_n[u_n({\bf x}_i)v_n^*({\bf x}_i+{\bf \delta})
+v_n^*({\bf x}_i)u_n({\bf x}_i  \nonumber \\
&&+{\bf \delta})]\tanh{E_n\over 2k_BT} ,
\label{DeltaV}
\end{eqnarray}
and the local inhomogeneous hole density is given by
\begin{eqnarray}
p({\bf x}_i)=1-2\sum_n[|u_n({\bf x}_i)|^2f_n+|v_n({\bf x}_i)|^2(1-f_n)],
\label{density}
\end{eqnarray}
where $f_n$ is the Fermi function. We stop the self-consistent calculations
only when all $p({\bf x}_i)\equiv p(i)$ converges to the appropriate
time CH density map like those shown in Fig.(\ref{MapVu}) or 
in Fig.(\ref{pMapDdw}).

Typical solutions can be visualized in Fig.(\ref{pMapDdw})
where we show the local density $p(i)$ on a square lattice of 28x28 sites 
where the  BdG calculations were made
and the 3D map of the low temperature amplitude of the $d_{x^2-y^2}$
order parameter $\Delta_d(i,T\approx0K)[cos(k_x)-cos(k_y)]$.
The pairing potential $V_{gb}(t)$ in unities of $eV$ is parametrized to 
yield the average local density of states (LDOS) gaps measured by 
low temperature STM  on the optimal doping $p=0.16$
$La_{2-p}Sr_pCuO_{2}$ \cite{Kato}, namely,
$\Delta_d(T\approx 0)\approx 7-12$meV. Starting with $V_{gb}(t=2000ts)=0.95t$, 
and analyzing 
the increase of  $V_{gb}(t)$ as those shown in Fig.(\ref{EV6200b}),
we were able to connect the time of X-ray exposure to
the superconducting amplitudes.

\begin{figure}[ht]  
\begin{center}
     \centerline{\includegraphics[width=9.0cm]{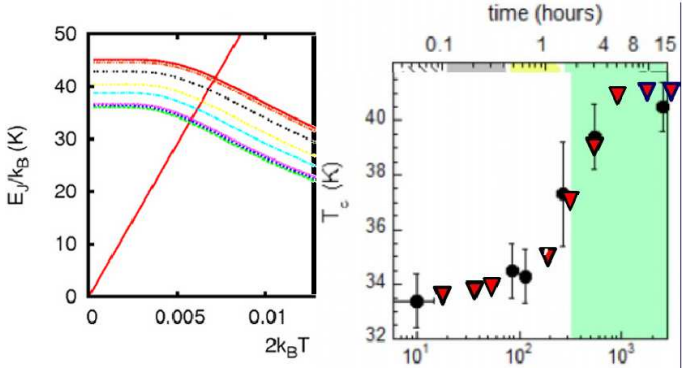}}
\caption{(color online) Left panel, the thermal energy
$k_BT$ and the Josephson coupling among superconducting grains $E_J(t,T)$
(as given by Eq.(\ref{EJ}))
for some given time of irradiation $t$ as function of T. The intersections
give the calculated results for $T_c(t)$ plotted
with inverted triangles on the top of the experimental results in 
the right panel.}
\label{EJTc}
 \end{center}
\end{figure}

With these values of the potential and the corresponding 
$\Delta_d(i,T\approx 0)$, we find that $\Delta_d(i,T)\rightarrow 0$ 
at a single temperature $T^*\approx96$K, which
is much larger than typical values of $T_c(p=0.16)$ for $La_2CuO_{4+y}$,
but in agreement with the pseudogap phase measured by STM\cite{Gomes},
or the Nernst signal on LSCO\cite{Ong,DDias07}. 
To obtain the measured
values of $T_c$, we notice that the EPS transition, 
with the free energy walls and wells, 
makes the structure of the system similar to granular 
superconductors\cite{Merchant}. In this way, it is likely that the
superconducting transition occurs in two steps\cite{Mello11,Mello12}: first by 
intra-grain superconductivity and than by Josephson coupling
with phase locking at a lower temperature. 
This two steps approach 
is also supported by the two
energy scales found in most cuprates\cite{Damascelli,LeTacon}, and
also in the two different regimes of critical fluctuations
in $YBa_2Cu_3O_7$ single crystals\cite{Pureur}.

 Granular superconductors are often modeled  as networks
of Josephson junctions (weak links) connecting the grains, i.e., the
Josephson network\cite{Ebner}. In the case of an order parameter
with $d_{x^2-y^2}$ symmetry,
the direction of the tunnel matrix element  across a junction is
essential and it was analyzed in detail\cite{Sigrist,Barash,Bruder}. 
In the special relative orientation for which the  d-wave gap
nodes in both side of the junction are parallel, the critical current behaves
mostly like an s-wave superconductor and its temperature dependence
is qualitatively the same (see, for instance, Fig.3a of Bruder 
et al\cite{Bruder}).
In the EPS considered here with the electronic 
grains in a $La_2CuO_{4+y}$ single crystal,
the superconducting order parameter in each grain has the same 
orientation in real space, with the nodes along the crystal
axis. Then the Josephson coupling is determined by
tunneling from a gap node in one grain into the corresponding  
node of another grain\cite{Sigrist,Barash,Bruder}.

Based on this similar behavior and on this special orientation, we use 
the analytical expression of the Josephson coupling energy $E_J$
to two s-wave superconductors\cite{AB}  

\begin{eqnarray} E_J(t,T) \approx \frac{\pi h\Delta^{av}_d(t,T)}{4 e^2 R_n}
tanh(\frac{\Delta^{av}_d(t,T)}{2K_BT}),
\label{EJ} 
\end{eqnarray} 
as an approximation
to obtain the relative changes of $E_J$ with the irradiation time.

In this way the increase of the potential
by the phase separation process enters in this equation by the values
of $\Delta^{av}_d(T,t)$.
Here $\Delta^{av}_d(T,t)\equiv \sum_i^N \Delta_d(T,i,t)/N$,
since the amplitude $\Delta_d(T,i,t)$ varies with the position $i$ within
the crystal, as shown in the right panel of Fig.(\ref{pMapDdw}). 
$R_n$ is the normal resistance of the $La_2CuO_{4+y}$
compound, which we assume to be independent of the time
as inferred from the data of Poccia et al\cite{Poccia}. It
is also proportional to the planar resistivity $\rho_{ab}$ 
measurements\cite{Takagi} on the
$La_{2-p}Sr_pCuO_2$ series. In Fig.(\ref{EJTc}),
the Josephson coupling $E_J(p,T)$ is plotted together with the
thermal energy $K_BT$. The intersections yield $T_c(t)$. 

As discussed in connection with the free energy of Eq.(\ref{FE}) and
Fig.(\ref{EV6200b}), the relative values of the grain boundary potential wall
$V_{gb}(t)$ are easy to estimate, but we do not know their absolute values.
As already mentioned, we use $V_{gb}(t=0.2h)=0.95t\approx 0.14$eV that matches the
1st point of Poccia et al\cite{Poccia}, namely, $T_c(t=0.2h)=33.6$K.
All the others $T_c(t)$ results for larger $t$, 
shown in Fig.(\ref{EJTc}), follow by the small but steady
variations in $V_{gb}(t)$ shown in the inset of Fig.(\ref{EV6200b}). 
In other words, we use $V_{gb}(t=0.2h)$ as
an adjustable parameter to obtain
$T_c(t=0.2h)$, but all the others seven points shown in the
right panel of Fig.(\ref{EJTc}) are parameter free, i.e., they follow from 
the increase of $V_{gb}(t)$ as shown in Fig.(\ref{EV6200b}).
The steep increase of $T_c(t)$ during the consolidation
of the stable phase between $t\approx0.3-5h$ agrees well with the
data as shown in Fig.(\ref{EJTc}), indicating that our triple approach
(CH simulation of $p(i)$, the local $\Delta_d(i,t)$ and the Josephson
estimation of $T_c(t)$ is appropriate to describe the $La_2CuO_{4+y}$ 
properties.

\section{Conclusions}

We have provided a complete description of a novel set
of complex experiments relating the time of sample preparation 
with the superconducting temperature $T_c(t)$.
The ensemble of calculations reported here demonstrated that the 
local variations in the free energy developed during the phase
separation process is an essential ingredient of the superconducting interaction.
These findings reveal a virtually unexplored line of research
on how systematic variations of sample preparation may affect
the superconductors properties.

I gratefully acknowledge partial financial aid from Brazilian
agencies FAPERJ and CNPq. 

\end{document}